\documentclass[showpacs,preprintnumbers,amsmath,amssymb]{revtex4}
\usepackage[pdftex]{graphicx}
\usepackage{graphicx}
\usepackage{subfigure}
\usepackage{longtable, tabu}

\begin{document}

\title{Validation of a track repeating algorithm for intensity modulated Carbon therapy with GEANT4}
\author{Qianxia Wang$^{1, 2}$}
\email{qw14@rice.edu}
\author{Antony Adair$^{1, 2}$}
\author{Yu Deng$^{3}$}
\author{Hongliang Chen$^{3}$}
\author{Michael Moyers$^{3}$}
\author{James Lin$^{3}$}
\author{Pablo Yepes$^{1,2}$}
\email{yepes@rice.edu}
\affiliation{$^1$Department of Physics and Astronomy, MS 315, Rice University, 6100 Main Street, Houston, TX 77005, USA}
\affiliation{$^2$Department of Radiation Physics, Unit 1420, The University of Texas MD Anderson Cancer, 1515 Holcombe
Blvd., Houston, TX 77030, USA}
\date{\today}

\begin{abstract}
The Fast Dose Calculator (FDC), a track repeating algorithm Monte Carlo method was initially
developed for proton therapy. The validation for proton therapy has been demonstrated in a previous work.
This method can be expanded to ion applications. Our purpose of this
paper is to validate the FDC for carbon therapy.
We compare the 3D dose distributions and dose-volume-histograms (DVH) for carbon calculated by FDC with a
full Monte Carlo method, GEANT 4. 19 patients in total will be discussed, including 3 patients of prostate, 5 of brain,
3 of head and neck, 4 of lung and 4 of spine. We use gamma-index technique to analyse dose distributions and we
do dosimetric analysis for DVH, a more direct and informative quantity for planning system assessment.
The FDC calculations of both quantities agree with GEANT4. The gamma-index passing rates of all patients discussed
in this paper are above 90\% with the criterion 1\%/1 mm, above 98\% with the criterion 2\%/2 mm and over 99.9\%
with the criterion 3\%/3 mm. The Root Mean Square (RMS) of percent difference
of dosimetric indices D$_{02}$, D$_{05}$, D$_{50}$, D$_{95}$ and D$_{98}$ are 0.75\%, 0.70\%, 0.79\%, 0.83\% and 0.76\%.
And all the difference are allowed for clinical use.
\end{abstract}


\maketitle

\section{Introduction}
Particle therapy (Wilson 1946, Amaldi 2005) is considered to have a greater potential to spare healthy tissue than 
traditional photon-therapy. It can deliver
dose to deep-seated or radioresistant tumors and cause less toxicity to the healthy tissue around the tumor (Castro $et$ $al$ 2004,
Schulz-Ertner et al 2007, Ohno 2013, Poludniowski et al 2015 ).
Thus in the last few years the number of particle therapy facilities has significantly increased (PTCOG website), in spite of 
their cost and technological challenges (Newhauser $et$ $al$ 2015). 
Compared with proton therapy, carbon therapy produces narrower lateral penumbra, which allows to minimize  
damage healthy tissue in the proximity the tumor. Moreover, carbon ions
have higher relative biological effectiveness (RBE) than protons (Kraft 2000). This feature causes more DNA double strand breaks
and lead to more non-repairable damage to tumor cells. Even though the cost of carbon therapy is 2-3 times more than
proton therapy, these avantages has boosted its clinical use. Moreover, good clinical results with carbon therapy have been
reported (Schulz-Ertner $et$ $al$ 2004, Tsujii $et$ $al$ 2004).

An essential component of any particle therapy treatment planning system is the dose calculation engine. 
Traditionally dose calculations was carried out with Pencil Beam Algorithms (PBS) (Petti 1992, Russell $et$ $al$ 1995, Hong $et$ $al$
1996, Deasy 1998, Schneider $et$ $al$ 1998, Schaffner $et$ $al$ 1999, Szymanowski and Oelfke 2002, Taylor $et$ $al$ 2017), due to
their calculation speed. However, it has been shown that Monte Carlo algorithms provide higher accuracy, especially
in areas with large homogeneities (Taylor $et$ $al$ 2017). Traditional Monte Carlo code require calculations times orders of magnitude
larger than PBS algorithms. However, in the last few years a variety of faster Monte Carlos have been developed
(Yepes $et$ $al$ 2009a , b, Dallas MC, Mayo MC, whatever else we can found). Among them, the only fast Monte Carlo for ion therapy is (Mayo).

Among the fast Monte Carlos, the Fast Dose Calcuator (FDC), a track-repeating Monte Carlo algorithm for protons was developed by 
Yepes et al. (Yepes $et$ $al$ 2009a , b), which can increase the calculation speed with respect to traditional MC by few orders
of magnitude. FDC was validated versus full Monte Carlo, GEANT4, for proton therapy in (Yepes $et$ $al$ 2016)。
In this work, we report in the extension of FDC to ion therapy and its validation.

\section {Methods}

\section {FDC Extension to Ions}
A stand-alone code, referred to as GEANT4, based on GEANT4 version 10.1.0 (Agostinelli $et$ $al$ 2003, Allison $et$ $al$ 2006), 
with the physics list FTF\_BERT was used for two purposes. 
Firstly it was utilized to generate the database of trajectories of
carbon ions, C$^{12}$, in water that was used as an input for FDC. Secondly it was employed to generate the reference dose distributions
for treatment plans for validation. 

The database of C$^{12}$ trajectories in water was generated by simulating 10K carbon ions with an energy of 5200 MeV 
impinging on a water phantom with the dimension of 510x510x2500 mm$^3$. For each impinging carbon, all the particles
(ions, protons, neutrons, electrons, and gammas) produced from it were stored. In addition, all the steps of the original particle
and of all its daughters were recorded in the database, along with the energy loss, the length, and direction for each step.

In addition to the trajectory database, parameters to scale the step length for different particles and materials were 
calculated and stored in a parameter repository. Similarly to the proton case, we considered a list
of 49 biological and other materials commonly encountered in radiation therapy (lucite, brass, etc). For each material and
a charged particle (ions, protons, and electrons), a table of the Relative Stopping Power (RSP) was stored 
as a function of particle kinetic energy in 1 MeV steps. 
RSP is defined as the stopping power of the material relative to water. The stopping power 
was obtained from the method ComputerTotalDEDX from G4EmCalculator for each particle, material and particle energy.

As for protons (Yepes $et$ $al$ 2016), the particle scaling parameters for scattering angles were obtained by taking the ratio of the
scattering angle in the material relative to water. However, this was implemented as function of particle energy in 1 MeV steps,
while in previous versions of the algorithm ratios were averaged over particle energies.  
Scattering angles of particles through a uniform slab of materials of thickness 0.02 g/cm$^2$ were calculated with 
the Moliere approximation, as implemented by Lynch and Dahl (1991). 

The basic track-repeating principle in FDC remains as in the proton case (Yepes 2009 a, 2009 b). 
However, the algorithm was updated to handle 
ions by utilizing the extended parameter repository with the length and angle scaling parameters
for ions produced in carbon collisions, as explained in the previous section. 

\section {Patient Cohort}

We selected 19 patients from different clinical sites treated at the University of Texas MD Anderson Cancer
Center (MDACC) with Intensity Modulated Proton Therapy (IMPT). The five clinical sites include prostate, brain,
head \& neck, lung and spine. For each type, 2 to 5 patients were selected for the study in this paper. 
The retrospective planning or dose calculations studies are conducted within purview of a generic protocol
approved by an MD Anderson Internal Review Board.

Since we did not have clinical ion plans available to us, we started with clinically used
proton plans, and  converted them into carbon treatment plans. Such conversion was achieved
by replacing for each energy layer the proton phase-space files describing the proton beam with carbon phase-space files, where protons of a given range were replaced with
carbons ions with the same range. Obviously, the dose distributions for the proton and carbon plans are not 
exactly the same, because of the different properties of protons and carbon ions. For example, carbon Bragg peaks
are sharper than those for protons, and they have a forward tail due to carbon fragmentation. In spite, of such 
differences, the resulting plans are meaningful for our comparison between FDC and GEANT4.

The target volume, prescribed dose, total voxel numbers, voxel size and maximum and minimum energies used in making plan are presented in Table \ref{table1}.

\begin{table}
\newcommand{\tabincell}[2]{\begin{tabular}{@{}#1@{}}#2\end{tabular}}
\begin{tabular}{|c | c | c | c |c| c |c |c |}
\hline 
Type      &Index & \tabincell{c}{Target Volumn\\ (cm$^3$)} &   \tabincell{c}{Prescr. Dose\\ (Gy)}&    Voxel \#   & \tabincell{c}{Voxel size\\ (mm$^3$)} & \tabincell{c}{Min Energy\\ (MeV/n)}  &   \tabincell{c}{Max Energy\\ (MeV/n)}\\ 
\hline
Prostate  & 1    &   21          &  65               &   17,796,597  &1.95$\times$1.95$\times$1.0                 &  264       & 361.26 \\
          & 2    &   31                &  22.0            &  21,294,338    &1.95$\times$1.95$\times$1.0                   &  267.7     &  381.9 \\
          & 3    &   18                &  38               &   11,408,683    &1.95$\times$1.95$\times$1.25                   &  303       & 375.5\\
\hline                                                                                             
Brain     & 1    &   39             &  54               &    12,424,230  &1.56$\times$1.56$\times$1.25                   &  137       & 270.5 \\
          & 2    &   63                &  54               &    11,195,197   &1.56$\times$1.56$\times$1.25                   &  135       &  294.5 \\
          & 3    &    11               &  55.4           &   11,967,150    &1.95$\times$1.95$\times$1.25                   &  173       &  250.5 \\
          & 4    &    24               &50.0             &   9,564,310      & 1.95$\times$1.95$\times$1.25                  &  169       &   278.5 \\
          & 5    &    63               &30.6             &    9,026,964    &1.95$\times$1.95$\times$1.25                &  135       &  247  \\
\hline                                                                                              
H \& N&  1   &    24               &70.0            &   6,482,515    &1.95$\times$1.95$\times$2.5                   &  270.5     & 346.5  \\
           & 2    &     7              & 66.0            & 18,624,294    &1.95$\times$1.95$\times$1.0                &  182.5     & 387.5  \\                                                                             
           & 3    &     14            & 66.0            &  22,667,190   &1.95$\times$1.95$\times$1.25                 &  219.5     &   286.5 \\
\hline                                                                                              
Lung  & 1    &    36              & 70.0            & 6,401,252    &1.95$\times$1.95$\times$2.5                &  158       &  298.5  \\
          &  2   &    33               &    66.0        &   10,125,024  &1.95$\times$1.95$\times$2.5                   &  181       &  250.5  \\
          & 3    &    117              &    63           &  5,101,360     &2.07$\times$2.07$\times$2.5                   &  204       & 377  \\
          &  4   &    125              &    66           &  4,270,560     &2.34$\times$2.34$\times$2.5                   &  274.5     &   387.5  \\
\hline                                                                                           
spine  &  1   &   138                  &    9.0      &  23,230,350   &1.95$\times$1.95$\times$2.0                &  237       &  316.5  \\
          &  2   &     48              &    45.5        &  6,794,229    &1.95$\times$1.95$\times$2.5                   &  209.5     &  332.5   \\
          &  3   &     333             &    70          &  10,505,404   &1.95$\times$1.95$\times$2.5                  &  135       &   298   \\
          &  4   &     36              &    50           &  11,393,369   &1.95$\times$1.95$\times$2.5                   &  215.5     &  312.5  \\
\hline
\end{tabular}   
\caption{Calculation details for all patients, which include target volume, prescribed dose, total number of voxel and minimum and maximum energies}
\label{table1}       
\end{table}

\begin{table}
\begin{tabular}{|c | c | c | c | c|c|c|c|c|c |c |c |}
\hline 
Type    &Index&GEANT 4 $\sigma$&FDC $\sigma$&P$_{11}$ (\%)& P$_{22}$ (\%)&P$_{33}$ (\%)& D$_{02}$ (\%)&D$_{05}$ (\%)&D$_{50}$ (\%)   & D$_{95}$ (\%)  &D$_{98}$ (\%) \\ 
\hline
Prostate& 1    &    0.396       &    0.376   & 96.81   & 99.82         & 99.99      & 0.83  &0.66  &0.51  &0.72  &0.73 \\
        & 2    &    0.425       &    0.416   & 95.82        & 99.64         & 99.98      & 0.91  &0.92  &0.95  &1     &1.02 \\
        & 3    &    0.378       &    0.307   & 95.80        & 99.61         & 99.97      &1.44   &1.45  &1.69  & 1.58 &1.42 \\
\hline                                                                                                                        
Brain   & 1    &    0.320    &    0.171   & 99.70     & 100.             & 100.         &- 0.52   &-0.53  &0     & 0.67 &0.79  \\
        & 2    &    0.356       &    0.191   & 99.67        & 100.          & 100.       & -0.32  &-0.32  & 0    & 0.36 &0.37 \\
        & 3    &    0.314       &    0.181   & 99.92        & 100.          & 100.       & -0.10  &-0.20  &0.11  &0.23  &0.23 \\
        & 4    &    0.171       &     0.105  & 99.95        & 100.          & 100.       &0.22   &0.22  &0.12  &0.38  &0.13 \\
        & 5    &    0.417       &    0.236   & 99.2         & 99.98         & 100.       &-0.30   &- 0.33 &0.59  &0     &0    \\
\hline                                                                                                                         
H \& N  & 1    &    0.201   &     0.112  & 98.71       & 99.7          & 100.       &0.85   &0.77  &0.61  &0.66  &0.33 \\
        & 2    &    0.456       &     0.334  & 96.56        & 99.93         & 100.       &0.5    &0.59  &1.19  & 1.38 &1.01 \\
        & 3    &    0.423       &     0.181  & 99.35        & 99.99         & 100.       &0.58   &0.68  &0.46  &0.91  & -0.22 \\
\hline                                                                                                             
Lung    & 1    &    0.335    &     0.175  & 99.26     & 98.99         & 100.       &-0.41   &-0.41  &-0.09  &0.19   &0.29 \\
        & 2    &    0.273       &     0.294  & 95.22        & 99.67         & 99.99      &1.5    &0.86  &1.33  &1.38  &1.41 \\
        & 3    &    0.393       &     0.198  & 97.63        & 99.92         &  100.      &0.85   &0.67  &0.61  &0.66  &0.33\\
        & 4   &    0.364       &     0.176   & 90.28        & 99.60         & 99.98      &1.23   &1.25  &1.42  &1.43  &1.64 \\
\hline                                                                                                                        
spine   & 1   &    0.257    &     0.269   & 98.42     & 99.97         & 100.           &0      &0     &0.73  &0     &0    \\
        & 2     &    0.189       &     0.141   & 99.10        & 99.98         & 100.       &0.51   &0.52  &0.57  &0.16  &0.17 \\
        & 3     &    0.426       &     0.207   & 98.46        & 99.97         & 100.       &-0.42   &-0.34  &0.09  &0.09  &0.1   \\
        & 4   &    0.290       &     0.136   & 99.93        & 100.          & 100.          &-0.50   &-0.75  &-0.67  &-0.91  &-0.78   \\
\hline                                                                                                     
RMS&          &    0.35        &     0.24    & 97.91       & 99.83         & 100.         &0.75   &0.70  &0.79  &0.83  &0.76  \\
\hline                                                                                                                                             
\end{tabular}   
\caption{Summary of gamma-index passing rates and difference between the GEANT4 and FDC dosimetric indices for the target
volume for all patients. Three different criteria are used for gamma-index calculation: 1\%/1mm (P$_{11}$), 2\%/2mm
(P$_{22}$ and 3\%/3mm (P$_{33}$). D$_{02}$, D$_{05}$, D$_{50}$, D$_{95}$ and D$_{98}$ are the maximum dose covering
2\%, 5\%, 50\%, 95\% and 98\% of the target.Values in the last line
are root mean square of all patients. }  
\label{table 2}     
\end{table}

\section {FDC-GEANT4 Comparisons}

Like in previous studies (Yepes $et$ $al$ 2016), we validate FDC by comparing its dose distributions to those obtained with GEANT4, a 
full-fledged Monte Carlo validated against measurements 
and widely utilized in the hadron therapy research. Each selected treatment plan is processed with FDC and GEANT4. 
Both methods are provided with the Radiation Therapy plan, the CT images and the structure
in DICOM format from the clinical treatment planning system.

In Table \ref{table 2}, we also give the statistical uncertainty of each voxel for Geant4 and FDC, which are shown on the
3rd and 4th column, respectively. The statistical uncertainty is related to the target volumn and number of histories used in the calculation.
Higher number of histories will bring down the statistical uncertainty. And larger target volumn needs more number of histories to get lower
statistical uncertainty.

The statistical uncertainty $\sigma$ can be calculated in the following steps:
$$
\overline{D}=\frac{1}{N}\sum_{i=1}^{N} D_i
$$,
$$
\sigma=\sqrt{\frac{1}{N}\sum_{i=1}^{N} (D_i-\overline{D})}
$$
where $i$ is the index of the each voxel, $D_ i$ is the dose deposit for each particle in every step in a particular voxel,
and $\overline{D}$ is the mean dose deposit. Note that for those voxels with dose below 10\% of the maximum dose are ignored.
It requires extra storage space. So it is only kept for voxel energy,  then transformed into statistical uncertainty in Dose.

The number of histories is 30 M for each beam when performing FDC calculation.
Because the computing time for GEANT4 is usually about 7000 to 18000 times more than FDC,
we did not use the same number histories as FDC. The number of histories selected
for GEANT4 is to make sure the statistical uncertainty blow 0.4. For all patients discussed in
this paper, the number of histories is between 10M to 65M for each beam. 
As can be noted, by comparing the 6th column with the 5th one,
most of the statistical errors for FDC are lower than that for GEANT4 because
the number of histories used in FDC is higher than that in Geant4. 

We compare the Geant4 and FDC 3D dose distributions by calculating 
the gamma-index (Low D A, Harms W B, Mutic S and Purdy J A 1998 A technique for the quantitative evaluation
of dose distributions Med. Phys. 25 656), which considers both the
difference in dose value and spatial position. It is calculated for all voxels with
a dose larger than 10\% of the maximum dose. We used an algorithm 
that uses the distance-to-simplex approximation as described in Ju et al 2008 (Ju et al 2008) 
with the criteria 1\%/1 mm, 2\%/2 mm and 3\%/3 mm.
A passing rate is calculated defined as the percentage of voxels that are within the tolerance (gamma index smaller than one). 

In addition, we calculate Dose Volume Histograms (DVHs) with FDC and GEANT4 and analyze the difference. DVHs
are calculated for structures contoured by physicians. We find that the larger disagreements between FDC and GEANT4 are found
for target volumes (TV). We attribute this behavior to the fact that TV are more sensitive to statistical 
fluctuations than DVHs of Organs at Risk (OAR). Therefore we have concentrated our analysis on this more 
challenging case of TVs. In order to evaluate the difference in DVHs, 
we compared D02, D05, D50, D95, and D98, defined as the the maximum dose covering 2\%, 5\%, 50\%, 95\% and 98\% of 
volume of the considered structure respectively.

\section {Results and discussion}

\subsection{Dose distribution}

\begin{figure}[h]
\begin{center}$
\begin{array}{c}
\includegraphics[width=120 mm]{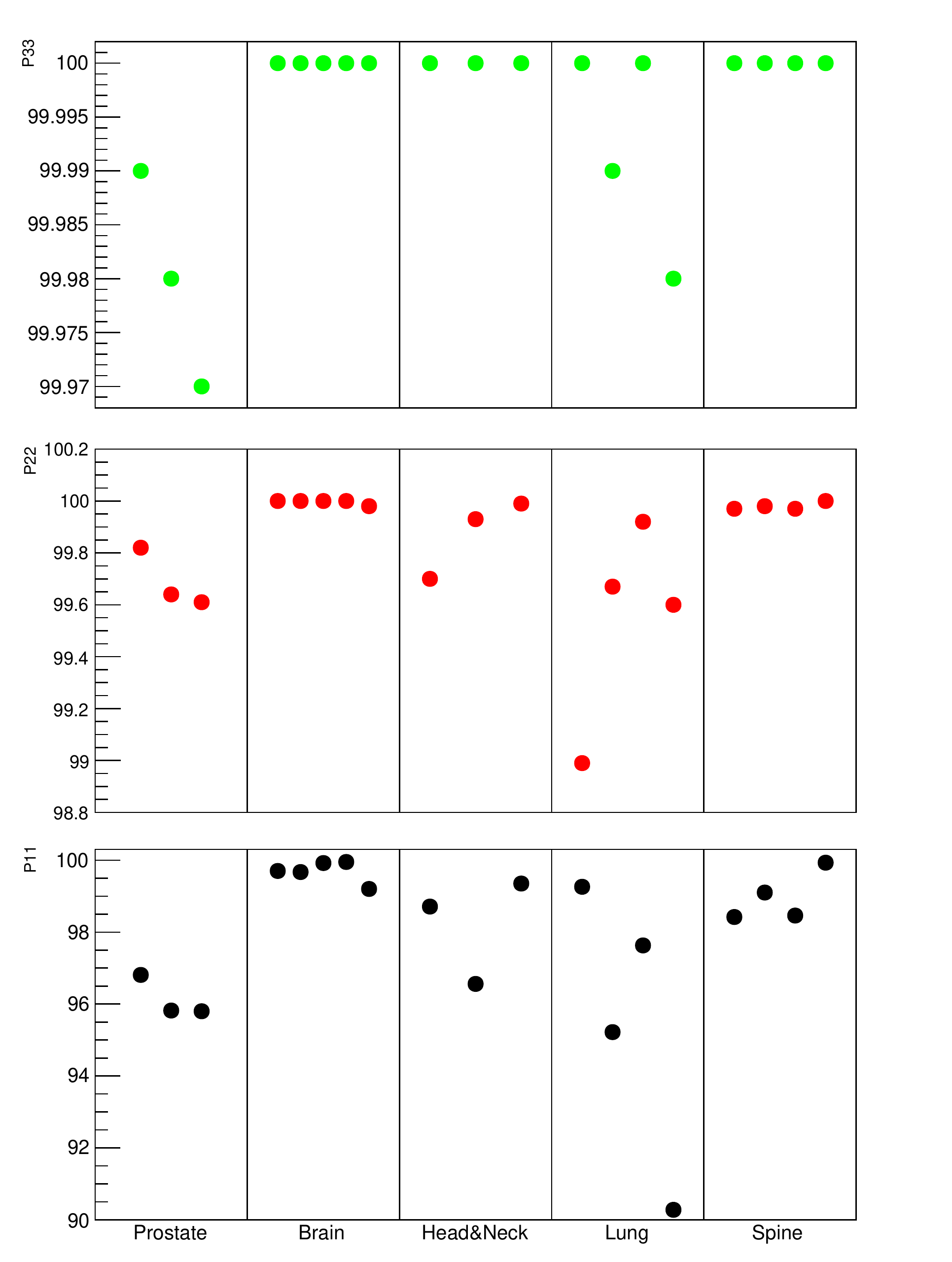} 
\end{array}$
\end{center}
  \caption{ Gamma-index passing rates for all patients of five different sites studied in this paper. From bottom to top, the criteria
used for Gamma-index calculation are 1 mm/1\% (P$_{11}$, black dots), 2 mm/2\% (P$_{22}$, red squares) and 3mm/3\% (P$_{33}$, green triangles).
 }
\label{Gamma-index}
\end{figure}

The gamma-index analysis is done for all patients with three different criteria, 1 mm/1\%, 2 mm/2\%
and 3 mm/3\%. The passing rates with different criteria for all patients were listed in the 5th, 6th and 7th column of Table II.
For all the patients, their passing rates are over 90.2\% for 1 mm/1\%, 98.9\%  for 2 mm/2\% and 99.9\% for 3 mm/3\%.
The RMSs are listed in the last line of Table II, which are 97.91, 99.83 and 100 for different criteria. The high passing rates demonstrate
that the FDC dose distributions agree well with that calculated by GEANT 4. These passing rates are visualized in Figure \ref{Gamma-index},
from bottom to top of which are 1 mm/1\% (P$_{11}$ black dots), 2 mm/2\% (P$_{22}$, red squares) and 3 mm/3\% (P$_{33}$, green triangles).
Figure \ref{Gamma-index} showed that the cohort of brain, H\&N and spine patients have higher gamma-index passing rates than
other three types. 

\begin{figure}[h]
\begin{center}$
\begin{array}{c}
\includegraphics[width=150mm]{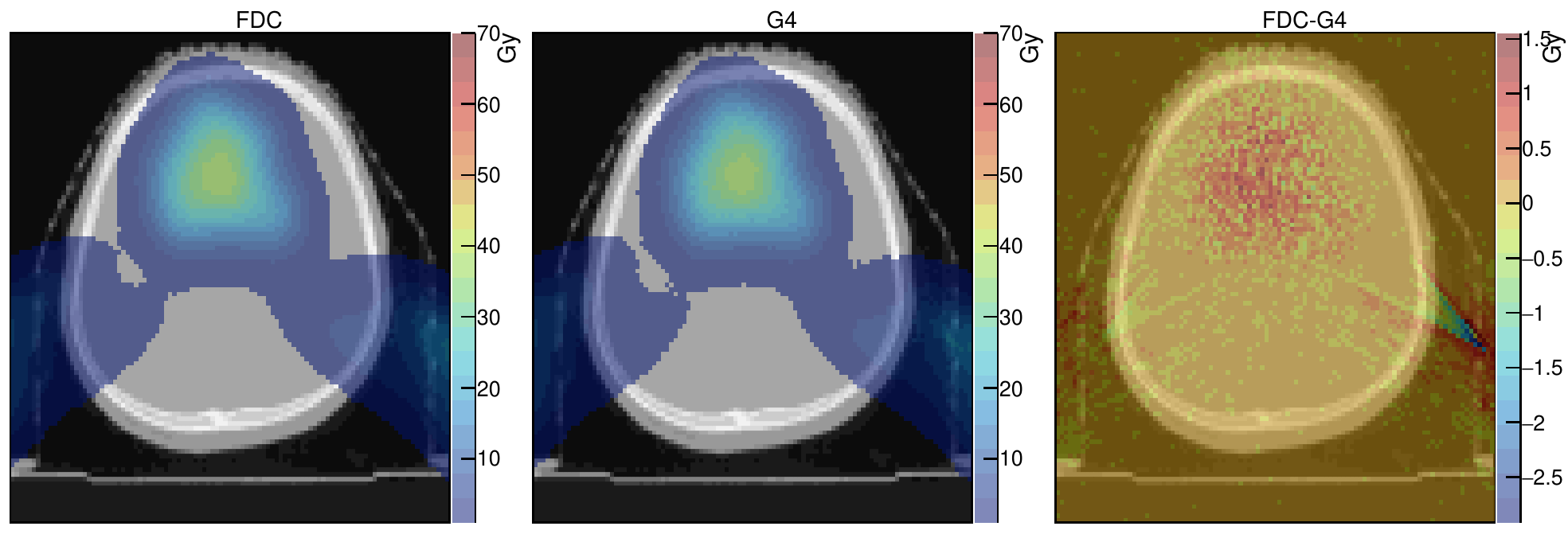} 
\end{array}$
\end{center}
  \caption{ Gamma-index passing rates for all patients of five different sites studied in this paper. From bottom to top, the criteria used
for gamma-index calculation are 1 mm/1\% (P$_{11}$, black dots), 2 mm/2\% (P$_{22}$, red squares) and 3mm/3\% (P$_{33}$, green triangles).
 }
\label{2D-dose}
\end{figure}

We select a head \& neck patient (Index 1) as an example to show 
the FDC-dose, GEANT4-dose and their difference distributions for a specific section in Figure \ref{2D-dose}. The
first plot in this figure is the dose distribution projected in the transverse plane with z=100 mm calculated
by FDC. The one in the middle is the dose projection in the same plane from GENAT4. The last panel is
the dose difference between two methods (FDC - GEANT4). As shown in the first two plots, the dose
distributions simulated by the two methods are similar to each other. Their differnce displayed in the last panel shows that
the maximum dose difference is of the order of 1 Gy, which confirms that the two methods agree with each other.

\begin{figure}[h]
\begin{center}$
\begin{array}{c}
\includegraphics[width=58 mm]{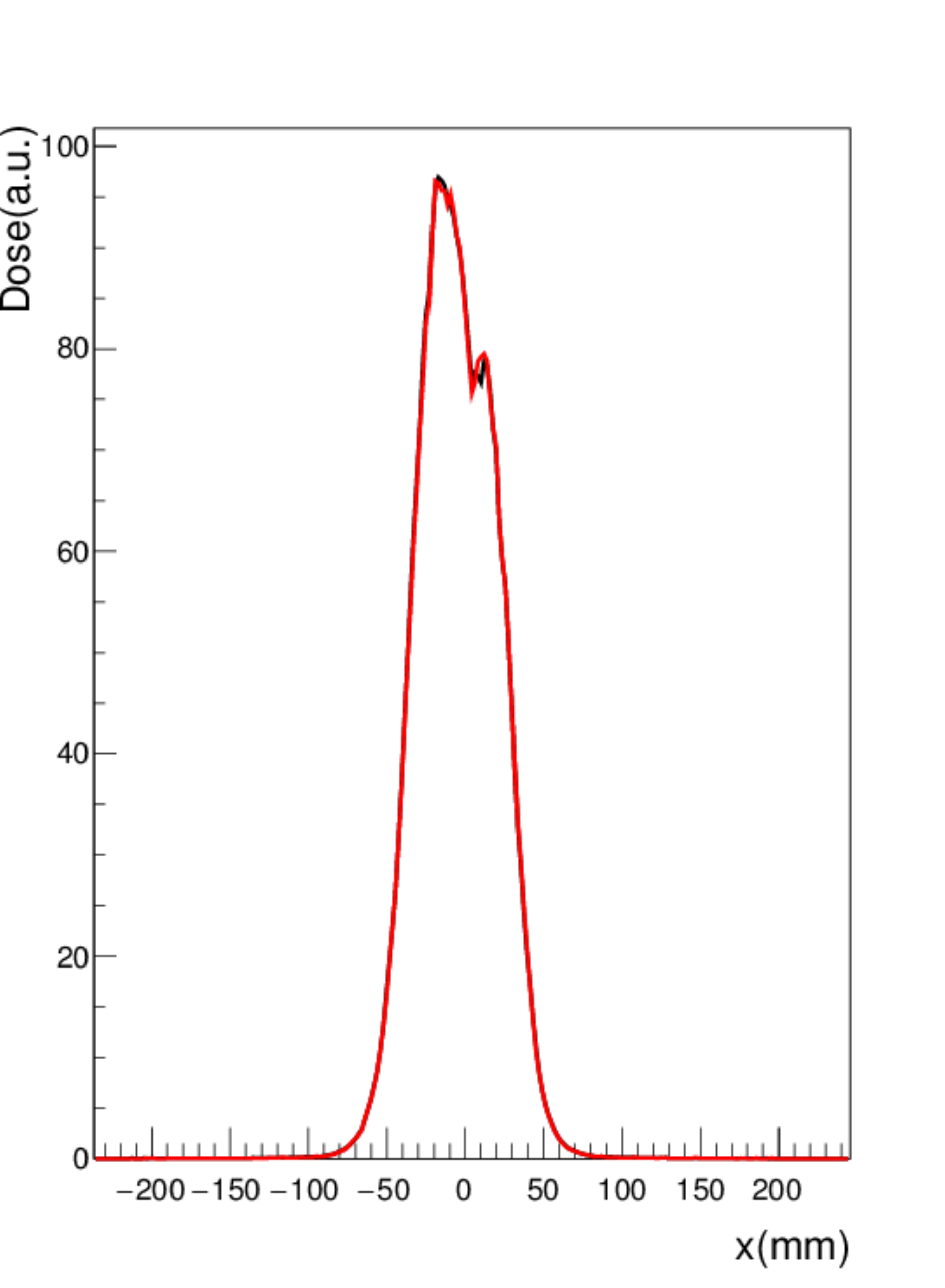} 
\includegraphics[width=58 mm]{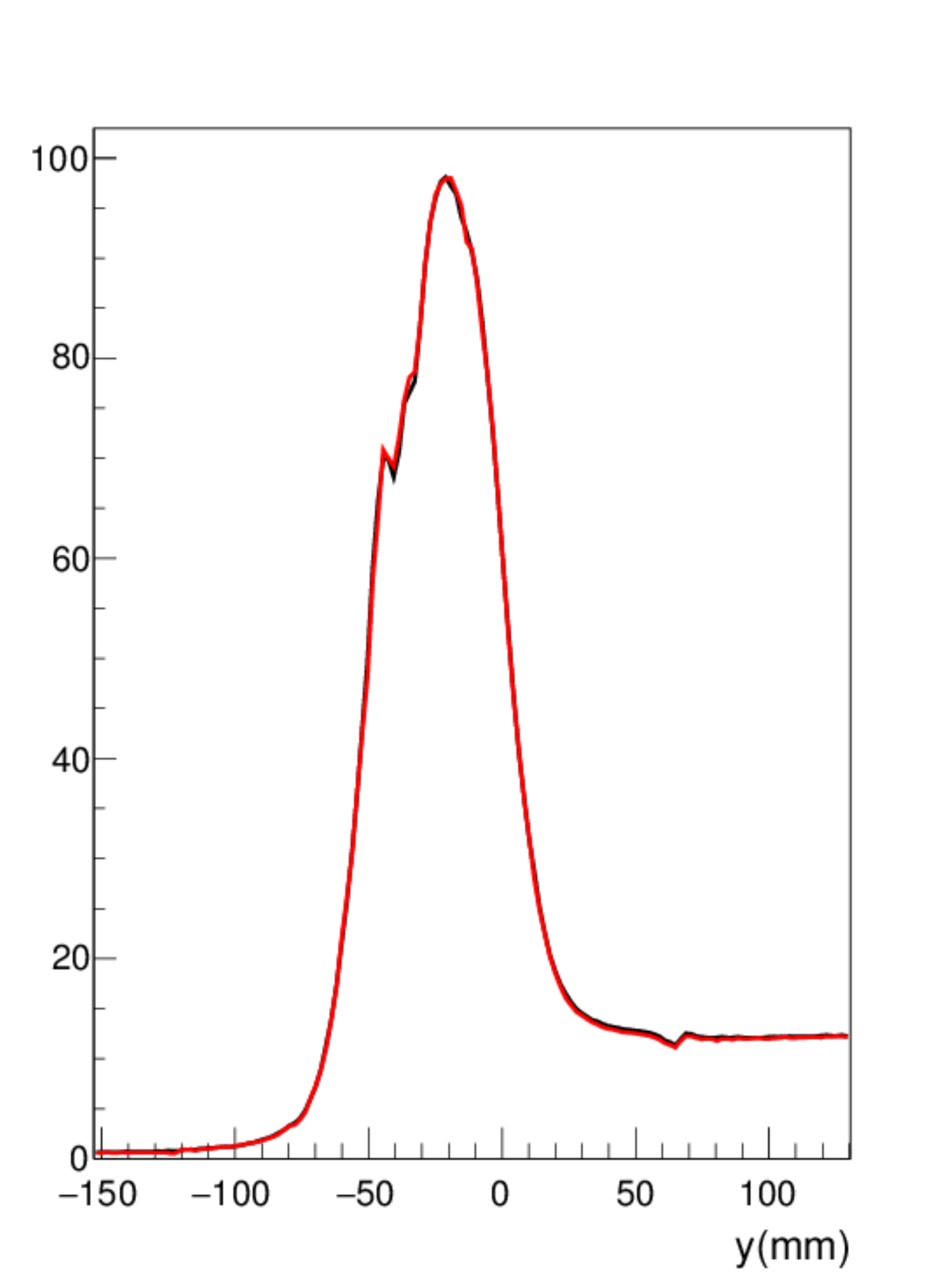}
\includegraphics[width=58 mm]{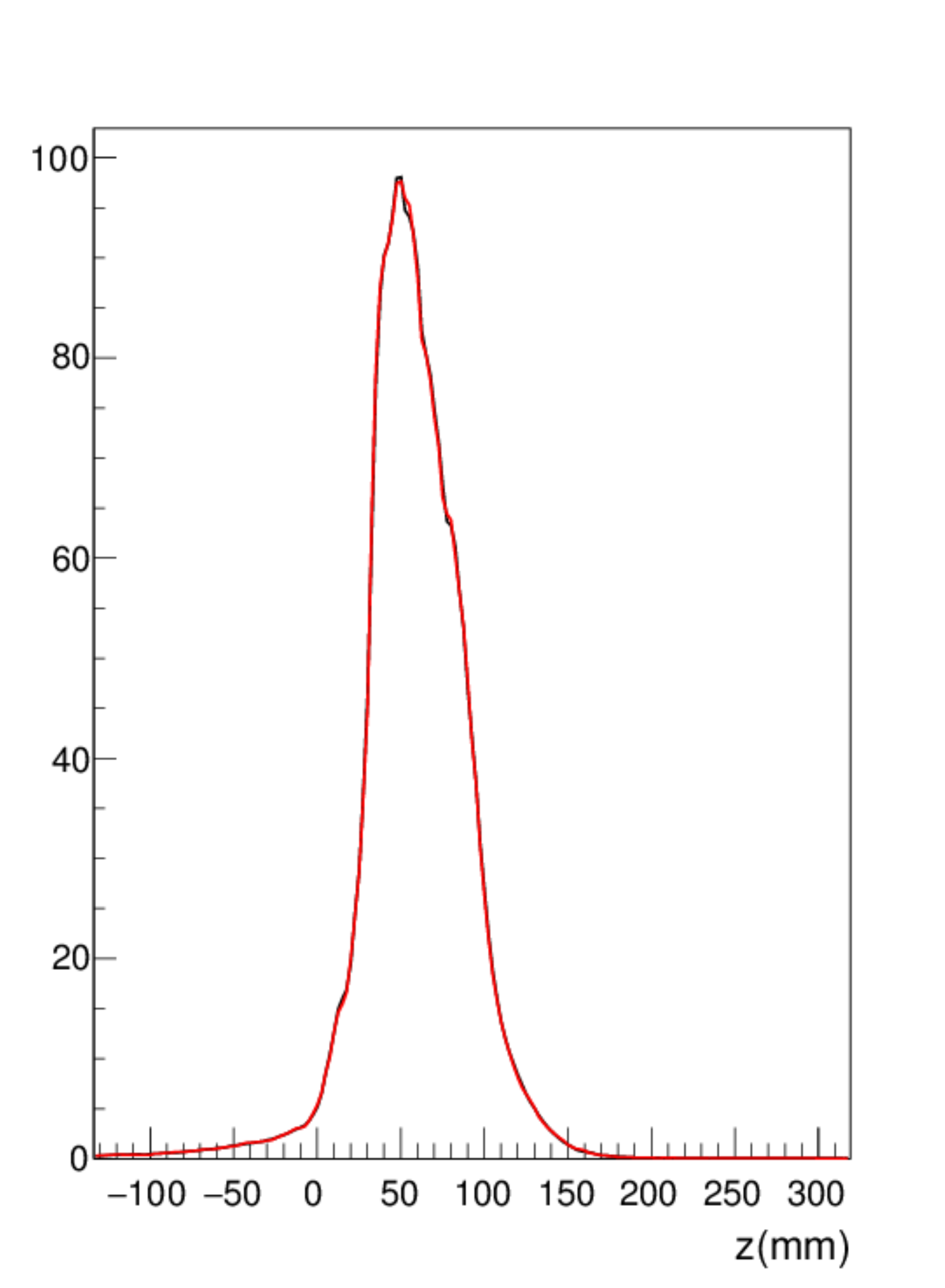}
\end{array}$
\end{center}
\caption{ Dose distribution projection in the direction superior to inferior (x), left to right (y) and anterior to posterior (z) for the same patient calculated by FDC (black line) and GEANT4 (red line)).
 }
\label{1D-dose}
\end{figure}

The dose distributions in x (lateral), y(anterior-posterior) and z (superior-inferior) projections for 
all the patients were also compared for a fast evaluation of the agreement. As an example, 
Figure \ref{1D-dose} displays the x, y and z projections for the same patient discussed above.
The curves of three projections calculated by the two methods coincide. 

\subsection{Dose-Volume-Histogram}

\begin{figure}[h]
\begin{center}$
\begin{array}{c}
\includegraphics[width=120 mm]{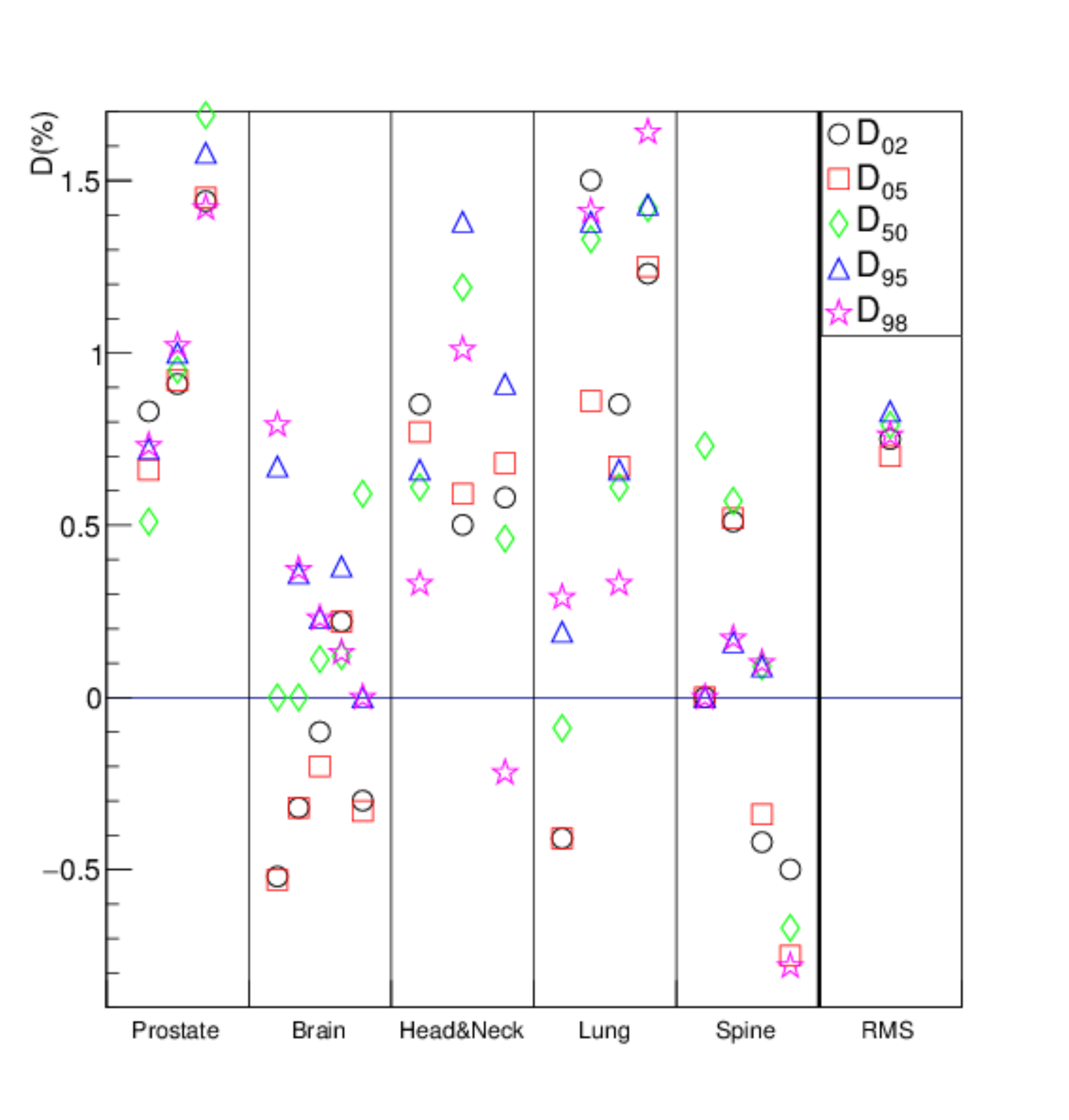} 
\end{array}$
\end{center}
  \caption{ Difference of five different indices of target volum DVH: D$_{02}$ 
D$_{05}$, D$_{50}$,  D$_{95}$, and D$_{98}$. Negative difference means FDC-DVH is small than GEANT4-DVH.
 }
\label{DVH}
\end{figure}

D$_{02}$, D$_{05}$, D$_{50}$, D$_{95}$ and D$_{98}$ for target volumn for all patients are calculated by GEANT4 and FDC.
For each index of DVH, the relative difference between the GEANT4 and FDC values was calculated and listed
in columns 8-12 of Table \ref{table 2}. The RMS of each index for all patients was calculated
and listed in the last line of Table \ref{table 2}. The difference in percentage
is below 1.5\% for D$_{02}$, 1.45\% for D$_{05}$, 1.69\% for D$_{50}$, 1.58\% for D$_{95}$ and 1.64 for D$_{98}$.
The RMSs of differences in D$_{02}$, D$_{05}$, D$_{50}$, D$_{95}$, D$_{98}$ calculated by two codes are 0.75\%, 0.70\%, 0.79\%,
0.83\% and 0.76\% respectively, which confirm that the DVH calculated by two codes agree well with each other.
All these values of difference between GEANT4-DVH and FDC-DVH for target are displayed in Figure \ref{DVH}.
The agreement for brain and spine patients is better than the other three types patients, which is consistent with the comparison of dose distribution.

\begin{figure}[h]
\begin{center}$
\begin{array}{c}
\includegraphics[width=150 mm]{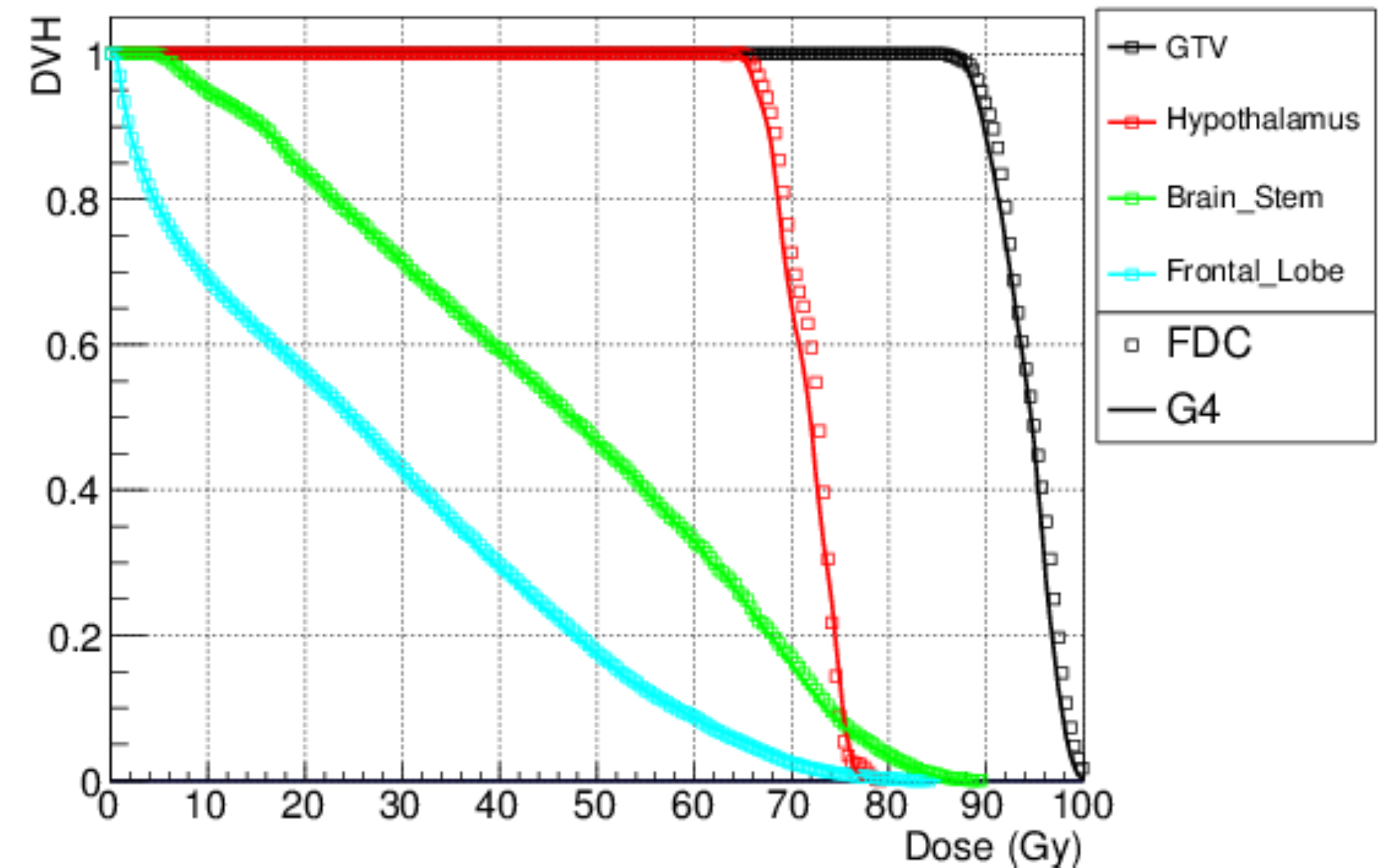} 
\end{array}$
\end{center}
  \caption{Dose-volume histograms of a head \&neck patient (Index 1) calculated
with FDC and GEANT4.
 }
\label{DVH-example}
\end{figure}

Figure \ref{DVH-example} displays DVHs of the same head \& neck patient mentioned above for four selected structures: GTV (black), Hypothalamus
(red), Brain$\_$Stem (green) and Frontal\_Lobe (light blue). The open squares and solid line are for FDC and GEANT4 calculations, repectively. The comparisons
show that the DVHs calculated by the two methods agree well with each other.  Especially,  the DVHs for Brain\_Stem and Frontal\_Lobe from the two
methods are virtually indistinguishable. 

\section{Conclusions}
In this paper, we used GEANT 4, a widely used full-fledged Monte Carlo code as the standard to verify the
accuracy of the Fast Dose Calculator (FDC) code in carbon patients calculation. We compared dose distributions
and  dose-volume-histograms (DVH) calculated by FDC and GEANT4.
The gamma-index passing rates with the criterion
2\%/2 mm are above 98.5\% for all patients, and the passing rates are above 99.9 for all patients if 3\%/3 mm
is used. For DVH, the Root Mean Square (RMS) for the difference of five selected slices (D$_{02}$, D$_{05}$, D$_{50}$, D$_{95}$, D$_{98}$)
calculated by FDC and GEANT4 are below 0.75\%, 0.70\%, 0.79\%, 0.83\% and 0.76\% respectively.
Therefore, the FDC accuracy amply satisfies the requirement for clinical use. 

\section{Acknowledgement}

\def\bibindent{1em}

\end{document}